# A sensitivity analysis of researchers' productivity rankings to the time of citation observation[1]

*Giovanni Abramo[a,b,*], Tindaro Cicero[b], Ciriaco Andrea D'Angelo[b]*

[a] Institute for System Analysis and Computer Science (IASI-CNR)
National Research Council of Italy

[b] Laboratory for Studies of Research and Technology Transfer
School of Engineering, Department of Management
University of Rome "Tor Vergata"

**Abstract**

In this work we investigate the sensitivity of individual researchers' productivity rankings to the time of citation observation. The analysis is based on observation of research products for the 2001-2003 triennium for all research staff of Italian universities in the hard sciences, with the year of citation observation varying from 2004 to 2008. The 2008 rankings list is assumed the most accurate, as citations have had the longest time to accumulate and thus represent the best possible proxy of impact. By comparing the rankings lists from each year against the 2008 benchmark we provide policy-makers and research organization managers a measure of trade-off between timeliness of evaluation execution and accuracy of performance rankings. The results show that with variation in the evaluation citation window there are variable rates of inaccuracy across the disciplines of researchers. The inaccuracy results negligible for Physics, Biology and Medicine.

# 1. Introduction

Continuous development in bibliometric indicators and techniques has made it possible to use bibliometrics to integrate or even totally substitute peer-review methods in national research evaluation exercises, at least for the hard sciences. In the United Kingdom, the previous peer-review Research Assessment Exercise series will be substituted in 2014 by the Research Excellence Framework (REF). The latter is an informed peer-review exercise, where the assessment outcomes will be a product of expert review informed by citation information and other quantitative indicators. In Italy there is a plan to substitute the peer-review Triennial Evaluation Exercise (VTR), first held in 2006, with a new Quality in Research Assessment (VQR). The new exercise can be considered a hybrid, as the panels of experts can choose from or use both methodologies for evaluating any particular output: i) citation analysis; and/or ii) peer-review by external experts. The Excellence in Research for Australia initiative (ERA), launched in 2010, is conducted through a pure bibliometric approach for the hard sciences: single research outputs are evaluated by a citation index referring to world and Australian benchmarks.

The pros and cons of peer-review and bibliometrics methods have been amply debated in the literature (Horrobin, 1990; Moxham and Anderson, 1992; MacRoberts and MacRoberts, 1996; Moed, 2002; van Raan, 2005; Pendlebury, 2009; Abramo and D'Angelo, 2011a). For evaluation of individual scientific products, the literature fails to decisively indicate whether one method is better than the other but demonstrates that there is certainly a correlation between the results from peer-review evaluation and those from purely bibliometric exercises (Franceschet and Costantini, 2011; Abramo et al., 2009; Aksnes and Taxt 2004; Oppenheim and Norris 2003; Rinia et al. 1998; Oppenheim 1997). The situation changes when evaluation turns from consideration of individual research products to ratings of individuals, research groups or entire institutions on a large scale. The huge costs and the long times of execution for peer-review force this type of evaluation to focus on a limited share of total output from each research institution. A number of negative consequences arise, among others: i) the final rankings are strongly dependent on the share of product evaluated; ii) the selection of products to submit to evaluation can be inefficient, due to both technical and social factors; and, most important iii) it is impossible to measure research productivity, which is the quintessential indicator of any production systems. Abramo and D'Angelo (2011a) have contrasted the peer-review and bibliometrics approaches in the Italian VTR and conclude that, for the hard sciences, the bibliometric methodology is by far preferable to peer-review in terms of robustness, validity, functionality, time and costs.

While peer-review can be applied to any type of research product at any moment after its codification, bibliometric methods, being based on citation analysis, are applicable only to research products for which citations are available. Furthermore, citation counts must be observed at sufficient distance in time from the date of publication in order to be considered a reliable proxy of real impact of a publication. The first condition means that the field of application for bibliometrics is limited to the hard sciences. The second one gives rise to a potential conflict between the need for evaluations to be conducted as quickly as possible after the period of interest and the need for time to develop accuracy and robustness in the ranking lists of individuals, research groups and institutions.

In order to provide policy makers and research institution managers a measure of the trade-off between timeliness of execution and accuracy of performance rankings, the authors have undertaken two studies: a first preparatory study, regarding the sensitivity of a publication's impact measurement to the citation window length (Abramo et al., 2011a), and a second concerning the sensitivity of the institutions' performance rankings (Abramo et al., 2011b). The conclusions were: i) with the sole exception of Mathematics, a time lapse of two or three years between date of publication and citation observation appears a sufficient guarantee of robustness in impact indicators for single research products (greater time lag would offer greater accuracy, but with ever decreasing incremental effect); ii) for rankings of institutional productivity, it seems sufficient to count citations one year after the upper limit of a three-year production period to ensure acceptable accuracy. In this work we complete the picture, investigating the sensitivity of individual researchers' productivity rankings to the time of citation observation. For this purpose we calculate the productivity of individual researcher staff in the hard sciences in Italian universities for the triennium 2001-2003, with the year of observing citations varying from 2004 to 2008. Comparing the rankings lists from each year to those from the 2008 benchmark we are then able to observe the trade-off between accuracy and timeliness in measurement. The following section of the paper presents the methodological details and dataset for the analysis. Section 3 shows the results from the elaborations. The final section provides a synthesis of the significant results and the author's considerations on policy implications.

## 2. Methodology and dataset

Research activity is a production process in which the inputs consist of human, tangible (scientific instruments, materials, etc.) and intangible (accumulated knowledge, social networks, etc.) resources, and where outputs have a complex character of both tangible nature (publications, patents, conference presentations, databases, protocols, etc.) and intangible nature (tacit knowledge, consulting activity, etc.). The new-knowledge production function has therefore a multi-input and multi-output character. In previous works of ours, we have measured research productivity on a national scale through non-parametric techniques (Abramo et al., 2011c), and at the individual level along a number of dimensions of output (Abramo and D'Angelo, 2011b). In this work we are not interested in assessing the bibliometric productivity of individual researchers per se, rather in finding out how productivity ranking lists vary with variation of the time of observation of citations. For this purpose, we will use a diachronous-prospective citation based indicator (Glänzel, 2004; Burrell, 2001; 2002), i.e. an indicator whose value changes with the time of citations observation, while the publication period remains the same.

We will adopt then a few simplifications and assumptions. To compare productivity of individual researchers we consider a single output production function: more precisely we measure the value of output, i.e. the impact, of their research activities in a given period of time, from 2001 to 2003. As proxy of overall output per researcher in the hard sciences, we consider publications (articles, article reviews and conference proceedings) indexed in Web of Science (WoS). As proxy of the value of output we adopt the number of citations for the researcher's publications. Because the intensity of publications varies by field, we compare researchers within the same field and rank them on a percentile scale. In the Italian university system all research personnel are classified in one and only one field. In the hard sciences, there are 205 such fields (named scientific disciplinary sectors, SDSs[2]), grouped into nine disciplines (named university disciplinary areas, UDAs[3]). When measuring labor productivity, if there are differences in the production factors available to each scientist then one should normalize by them. Unfortunately relevant data are not available. However, assuming a uniform distribution of capital per research staff, is not far from reality in Italy. Here, the large part of financial resources is equally allocated by government to satisfy the needs of each university in function of its size. The potential greater availability of funds per staff unit in a university is thus due to its capacity to acquire such funds on a competitive basis. Greater output deriving from greater availability of funds is thus the result of merit and not of any other comparative advantages. Furthermore, it is not unlikely that researchers belonging to a particular scientific field may also publish outside that field. For this reason we standardize the citations for each publication accumulated at December 31 of each year from 2004 to 2008 with respect to the median[4] for the distribution of citations for all the Italian publications of the same year and the same Web of Science (WoS) subject category[5]. The productivity of a single researcher, named Scientific Strength (SS)[6], is given by:

$$SS = \sum_{i=1}^{N} \frac{c_i}{Me_i}$$

Where:

$c_i$ = citations received by publication $i$;

$Me_i$ = median of the distribution[7] of citations received for all Italian publications of the same year and subject category of publication $i$;

N = number of publications of the researcher in the period of observation.

We elaborate researchers' SS ranking lists for each SDS and for each year of the citation window from 2004 to 2008.

Data on research staff of each university and their SDS classification are extracted from the database on Italian university personnel, maintained by the Ministry for Universities and Research[8]. The bibliometric dataset used to measure productivity is extracted from the

---

[2] The complete list is accessible on http://attiministeriali.miur.it/UserFiles/115.htm

[3] Mathematics and computer sciences; physics; chemistry; earth sciences; biology; medicine; agricultural and veterinary sciences; civil engineering; industrial and information engineering.

[4] As frequently observed in literature (Lundberg, 2007), standardization of citations with respect to median value rather than to the average is justified by the fact that distribution of citations is highly skewed in almost all disciplines.

[5] The subject category of a publication corresponds to that of the journal where it is published. For publications in multidisciplinary journals the median is calculated as a weighted average of the standardized values for each subject category.

[6] SS is similar to the "crown indicator" of CWTS (Moed et al., 1985) and the "total field normalized citation score" of the Karolinska Institute (Rehn et al., 2007). The differences are: i) we standardize citations of single publications and not of scientific portfolio of researchers; ii) we standardize by the Italian median rather than the world average.

[7] Publications without citations are excluded from calculation of the median.

[8] http://cercauniversita.cineca.it/php5/docenti/cerca.php. Last accessed on December 7, 2011.

Italian Observatory of Public Research (ORP)[9], a database developed and maintained by the authors and derived under license from the Thomson Reuters WoS. Beginning from the raw data of the WoS, and applying a complex algorithm for reconciliation of the author's affiliation and disambiguation of the true identity of the authors, each publication (article, article review and conference proceeding) is attributed to the university scientist or scientists that produced it (D'Angelo et al., 2010).

To ensure the representativity of publications as proxy of the research output, the field of observation was limited to those SDSs where at least 50% of researchers produced at least one publication in the period 2001-2003. Furthermore, we excluded those SDSs with fewer than 10 members. We thus considered a total of 174 SDSs, with the dataset composed of 146,569 publications authored by a total of 20,634 academic scientists. The distribution of publications among the 174 SDSs and 9 UDAs is presented in Table 1.

| UDA | n° of research staff | n° of publications | n° of SDSs | n° of universities |
|---|---|---|---|---|
| Mathematics and computer sciences | 1,714 | 7,061 | 9 | 54 |
| Physics | 1,858 | 24,097 | 8 | 55 |
| Chemistry | 2,401 | 23,114 | 11 | 55 |
| Earth sciences | 706 | 2,824 | 12 | 44 |
| Biology | 3,377 | 21,978 | 19 | 57 |
| Medicine | 5,959 | 41,887 | 42 | 49 |
| Agricultural and veterinary sciences | 1,458 | 5,965 | 28 | 34 |
| Civil engineering | 585 | 2,195 | 7 | 39 |
| Industrial and information engineering | 2,576 | 17,448 | 38 | 54 |
| Total | 20,634 | 146,569 | 174 | 65 |

*Table 1: Number of Italian universities, research staff, SDSs and publications per UDA; data 2001-2003*

## 3. Results and analysis

In order to decide on the time lag to allow after the period of interest and prior to bibliometric evaluation of researcher productivity, the decision maker must be informed on the inaccuracy in rankings in function of time lapse. The inaccuracy can be measured along various dimensions, which can then be weighted according to the context. In our opinion, the main dimensions of inaccuracy to consider are: i) the number of researchers which experience substantial variation in ranking (preferably indicated as numbers penalized and numbers that received an advantage); ii) the overall average of shifts in rank position; and iii) the maximum shift in rank. The final choice by the decision-makers will be a compromise between the inaccuracy that they are willing to accept and the time that they are willing to wait prior to carrying out the evaluation: these factors will also depend on context. In the following, we elaborate SS ranking lists of researchers for each SDS and for each year of the citation window 2004 to 2008. The 2008 ranking list is the most accurate, as citations have had the longest time to accumulate, thus representing the best possible proxy of impact. We choose it as the benchmark to estimate inaccuracy occurring in rankings calculated in previous years. Naturally the 2008 ranking list will in turn be less accurate than lists elaborated in successive years, being based on larger citation windows, but it is legitimate to believe that an exercise at more than five years distance from the last year of the period of observation would be of little use toward the evaluation objectives.

---

[9] www.orp.researchvalue.it. Last accessed on December 7, 2011.

We proceed by first showing the variability of rankings relative to time of citation observation for a single generic SDS, then we compare the SDSs of a single UDA and finally we compare all the hard science UDAs, in order to detect potential differences across SDSs and UDAs. The sensitivity analyses are conducted first for percentile and then for quartile rankings.

**3.1 Volatility of productivity percentile rankings of researchers**

To compare SDSs with various numbers of members, the productivity rankings of researchers are expressed as percentile rounded to whole numbers (scale from 0-100). Assuming the 2008 evaluation as benchmark, Figure 1 shows the frequency distribution for shifts in rank of researchers active in the Geometry SDS, given variation in the year of counting citations. The choice of an example SDS from the Mathematics UDA is due to the characteristic behavior of this UDA to accumulate citations somewhat more slowly than others (Abramo et al., 2011a).

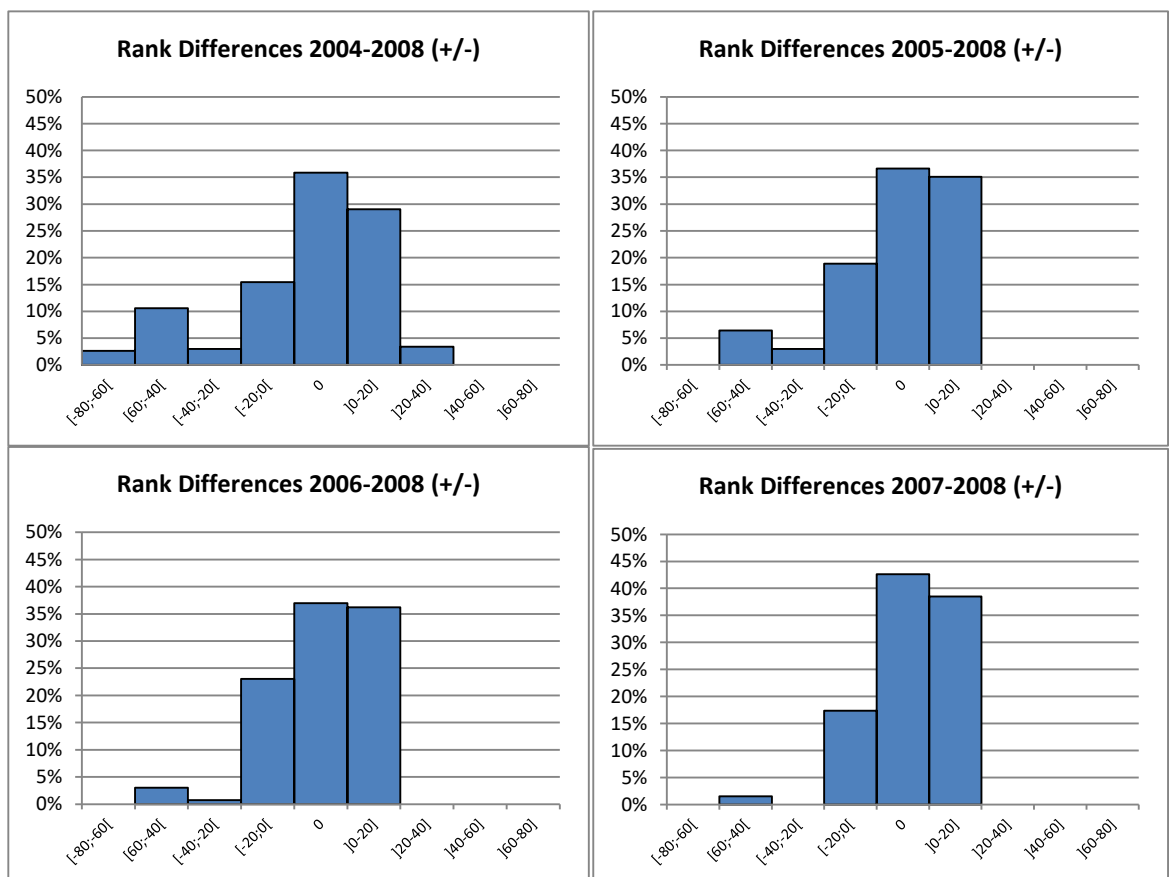

*Figure 1: Frequency distributions of differences between 2008 researchers' productivity ranking*

*(percentiles) and previous years in Geometry (265 researchers)*

The comparison between the 2004 and 2008 rankings lists shows a distribution of shifts that is skewed to the left and unimodal (mode: 0 shifts). We also observe that the positive shifts are totally concentrated in the two lower classes, while negatives shifts are dispersed over the four classes of shifts considered. In detail, just over 15% of observations are of 0 to -20 shifts in rank, while for the corresponding positive class (between 0 and +20s shifts), the percentage of cases reaches almost 30%. In successive years the frequencies of rank variation show distributions that are more concentrated in the central class (nil shift) with less dispersion among the remaining classes. In comparing 2007 to 2008 rankings, more than 40% of researchers show nil shift.

Table 2 shows the statistics of the Geometry SDS for distribution for absolute shifts in rank and positive and negative subgroups. Comparing to benchmark, the percentage of those with a change in their rank descends from 64.2% in 2004 to 57.4% in 2007. Separating the analyses for the individuals who experience positive and negative changes in rank, we observed that the percentage of those in the first group tends to slight increase until 2007, going from 32.5% to 38.5%, while there is a pronounced decrease in the second group, from 31.7% to 18.9%. The average value of shift for the two subgroups is notably different: for those who experience a negative shift, the average change in 2004 is -27.4 ranks, but it is only +9.8 for those on the positive side, compared to an overall average value of 11.9 for distribution of absolute shifts. The differences tend to reduce going towards 2007, but remain significant. In every year the maximum shift in rank is always negative: in 2004 there is a researcher with a -76 shift in rank, while in 2007 there is a shift of -48 positions.

However, the correlation value between the distributions of differences in rank is already quite high in 2004 (+0.884) and in 2005 the value reaches +0.947.

| Descriptive statistics | | 2004-2008 | 2005-2008 | 2006-2008 | 2007-2008 |
|---|---|---|---|---|---|
| % change/ total researchers | General | 64.2% | 63.4% | 63.0% | 57.7% |
| | Δ+ | 32.5% | 35.1% | 36.2% | 38.5% |
| | Δ- | 31.7% | 28.3% | 26.8% | 18.9% |
| Average change | General | 11.9 | 7.4 | 4.5 | 2.3 |
| | Δ+ | +9.8 | +6.8 | +4.6 | +2.5 |
| | Δ- | -27.4 | -17.6 | -10.4 | -7.1 |
| Median change | General | 5 | 3 | 2 | 1 |
| | Δ+ | +7 | +6 | +4 | +3 |
| | Δ- | -24 | -10 | -4 | -3 |
| Maximum | General | 76 | 58 | 48 | 48 |
| | Δ+ | +30 | +17 | +14 | +6 |
| | Δ- | -76 | -58 | -48 | -48 |
| Standard deviation | General | 17.5 | 11.7 | 8.4 | 5.6 |
| Spearman correlation | General | +0.884 | +0.947 | +0.973 | +0.989 |

*Table 2: Descriptive statistics of distributions of differences of researchers' productivity rankings (percentiles) in Geometry*

This analysis was repeated for all 174 SDSs. We present the additional example of the statistics for Experimental physics SDS (Physics UDA) which has a citation pattern much different from the SDSs of Mathematics: here there is a very rapid accumulating of citations (Table 3).

| Descriptive statistics | | 2004-2008 | 2005-2008 | 2006-2008 | 2007-2008 |
|---|---|---|---|---|---|
| % change/ total researchers | General | 84.5% | 81.0% | 76.7% | 65.4% |
| | Δ+ | 45.0% | 43.4% | 41.8% | 36.0% |
| | Δ- | 39.5% | 37.6% | 34.9% | 29.4% |
| Average change | General | 4.0 | 2.7 | 1.8 | 1.1 |
| | Δ+ | +4.3 | +3.0 | +2.2 | +1.6 |
| | Δ- | -5.2 | -3.6 | -2.7 | -1.9 |
| Median change | General | 3 | 2 | 1 | 1 |
| | Δ+ | +3 | +3 | +2 | +1 |
| | Δ- | -4 | -3 | -2 | -1 |
| Maximum | General | 25 | 15 | 10 | 7 |
| | Δ+ | +17 | +10 | +9 | +6 |
| | Δ- | -25 | -15 | -10 | -7 |
| Standard deviation | General | 4.0 | 2.5 | 1.8 | 1.2 |
| Spearman correlations | General | +0.982 | +0.992 | +0.996 | +0.998 |

*Table 3: Descriptive statistics of distributions of differences of researchers' productivity rankings (percentiles) in Experimental physics (789 researchers)*

The situation seems paradoxical in comparison to the Geometry SDS, in that the percentage of researchers who experience a change in rank is actually higher. In 2004, approximately 84% of researchers shift rank compared to benchmark, and in 2007 the percentage is still 65%. But from the average values we can see that these many changes in rank are actually minor fluctuations: in 2004, the average shift in rank for researchers is 4 positions and there are no substantial differences between the two subgroups of those who shift in positive and negative sense. The correlation between differences in rank is already very high in the first year of evaluation (+0.982) and tends almost to one by 2007.

We have now observed how two SDSs, from UDAs with different characteristics in citation patterns, show very different levels of inaccuracy in researcher evaluation with variation of the time for counting citations. However it is important to understand if there are fluctuations across SDSs that belong to a single UDA. Towards this, Table 4 presents descriptive statistics for all the SDSs in the Physics UDA. With the exception of FIS/06 (Physics for earth and atmospheric sciences) and FIS/08 (Didactics and history of Physics), where researchers show average shifts in 2004-2008 ranks of 9.0 and 7.4 places, in all other SDSs the average rank differences are quite limited, ranging from 3.7 in FIS/04 to 5.8 in FIS/07, and in all cases the coefficients of correlation show rapid convergence of the rankings.

| SDSs[10] | Descriptive Statistics | 2004-2008 | 2005-2008 | 2006-2008 | 2007-2008 |
|---|---|---|---|---|---|
| FIS/01 | Average | 4.0 | 2.7 | 1.8 | 1.1 |
|  | Maximum | -25 | -15 | -10 | -7 |
|  | Spearman | +0.982 | +0.992 | +0.996 | +0.998 |
| FIS/02 | Average | 4.9 | 3.4 | 2.4 | 1.3 |
|  | Maximum | -23 | +/-15 | -13 | -11 |
|  | Spearman | +0.976 | +0.988 | +0.994 | +0.998 |
| FIS/03 | Average | 5.0 | 3.0 | 2.1 | 1.3 |
|  | Maximum | -23 | +/-13 | -12 | +7 |
|  | Spearman | +0.973 | +0.990 | +0.995 | +0.998 |
| FIS/04 | Average | 3.7 | 2.7 | 1.9 | 1.5 |
|  | Maximum | -20 | -15 | +8 | -7 |
|  | Spearman | +0.985 | +0.992 | +0.996 | +0.998 |
| FIS/05 | Average | 4.2 | 2.9 | 1.9 | 1.3 |
|  | Maximum | -23 | -13 | -11 | -8 |
|  | Spearman | +0.980 | +0.990 | +0.996 | +0.998 |
| FIS/06 | Average | 9.0 | 5.0 | 3.0 | 2.3 |
|  | Maximum | -35 | +22 | -20 | -15 |
|  | Spearman | +0.912 | +0.971 | +0.982 | +0.992 |
| FIS/07 | Average | 5.8 | 4.0 | 2.9 | 1.8 |
|  | Maximum | -28 | -18 | -17 | -12 |
|  | Spearman | +0.965 | +0.983 | +0.990 | +0.995 |
| FIS/08 | Average | 7.4 | 6.5 | 4.1 | 2.1 |
|  | Maximum | -39 | -39 | -22 | -17 |
|  | Spearman | +0.936 | +0.952 | +0.978 | +0.989 |

*Table 4: Descriptive statistics of distributions of differences of researchers' productivity rankings (percentiles) in Physics by SDS*

For a general view of the differences in rankings between the 2004 and 2008 listings we calculated the productivity of researchers in each SDS. Figure 2 shows the accumulated frequencies curve for the Spearman correlation index of the two rankings and Figure 3 shows cumulative frequency of average shifts in each SDS. We see there is no SDS with Spearman index less than 0.5; in 92% of the SDSs the index is over 0.8 and in 2/3 of these (67.1% of the 174 total SDSs) it is greater than 0.9. The distributions of average values of differences in rankings (by percentile) also show substantial convergence of the 2004 evaluation to the 2008 results. In 90.2% of SDSs the average value of ranking shift is not more than 15 percentiles and in 76.4% of cases the shift is less than 12.

---

[10] FIS/01 = Experimental Physics; FIS/02 = Theoretical physics, mathematical models and methods; FIS/03 = Material physics; FIS/04 = Nuclear and subnuclear physics; FIS/05 = Astronomy and astrophysics; FIS/06 = Physics for earth and atmospheric sciences; FIS/07 = Applied physics (cultural heritage, environment, biology and medicine); FIS/08 = Didactics and history of physics.

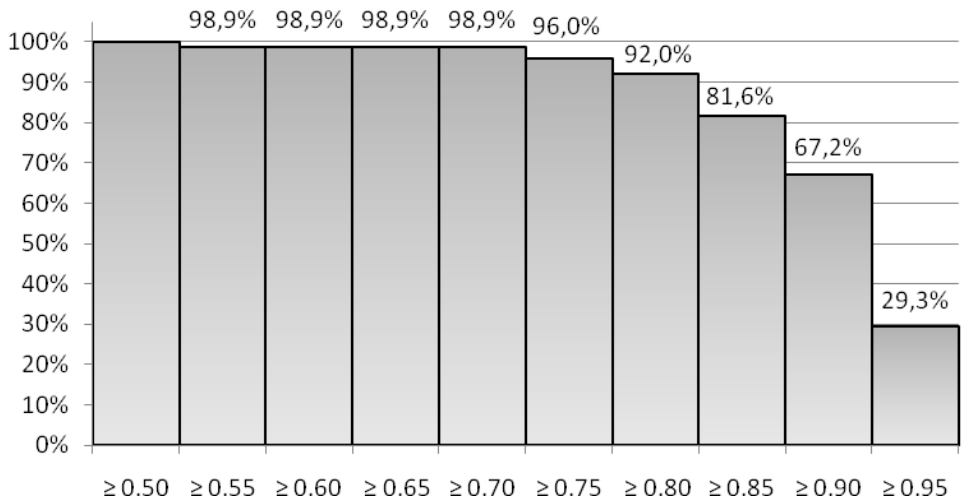

*Figure 2: Spearman correlation of researchers' productivity rankings 2004 and 2008; cumulative frequency for all 174 SDSs examined*

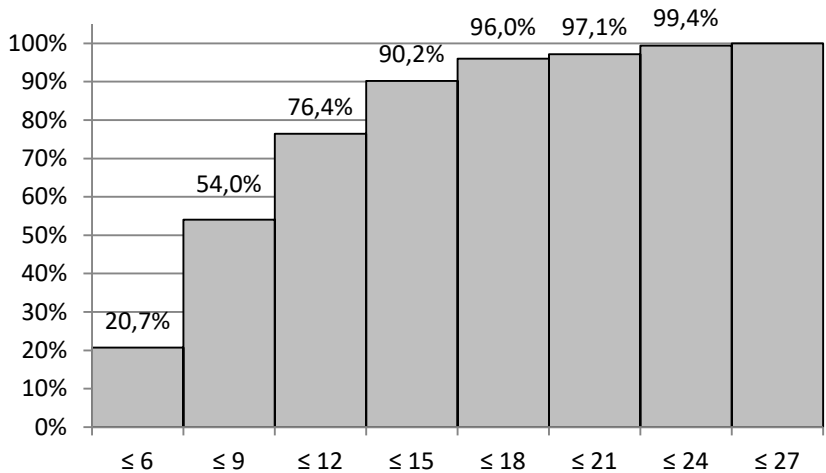

*Figure 3: Cumulative frequency of average differences of researchers' productivity rankings (percentiles) 2004 vs. 2008, for all 174 SDSs examined*

Finally, analyses were conducted to detect differences across UDAs, shown in the descriptive statistics of Table 5. For reasons of space, we present for each UDA only the SDSs that show the maximum and minimum average values of shift between evaluation in 2004 and 2008. The UDAs with the greatest average shift are Medicine and Industrial and information engineering. However these two UDAs show structural differences. In Medicine there is the case of a researcher (SDS MED/34, Physical and rehabilitation medicine) who shifts 100 positions of a maximum possible 100 and thus increases the average value of shift for 2004 to 22.7; but from 2005 onward the average value gets very low; at 5.0 for 2006, and 6.2 for 2007[11]. This trend suggests that the SDS features one or more outlier researchers and that the UDA is otherwise characterized by shifts that are much lower in average value. In Industrial and information engineering there is again an SDS (ING-IND/12, Mechanical and thermal measuring systems) with a researcher who makes the maximum shift of 100 positions, causing a high average value of 24.1 for 2004; but unlike the case of MED/34, the average shift in 2005 remains quite high (average = 15.0) and then decreases in a fairly linear manner until 2007.

Other UDAs characterized by SDSs with high variability in rankings are Mathematics and computer sciences, Earth sciences and Agricultural and veterinary sciences.

| UDAs | SDSs[12] | Index | 2004-2008 | 2005-2008 | 2006-2008 | 2007-2008 |
|---|---|---|---|---|---|---|
| Mathematics and computer sciences | MAT/01 | Average | 15.7 | 5.2 | 2.9 | 2.6 |
| | | Max | -50 | +20 | +10 | ±10 |
| | | Spearman | +0.779 | +0.976 | +0.990 | +0.989 |
| | MAT/07 | Average | 9.1 | 5.4 | 3.3 | 1.4 |
| | | Max | -76 | -31 | -21 | -21 |
| | | Spearman | +0.919 | +0.967 | +0.987 | +0.997 |
| Physics | FIS/06 | Average | 9.0 | 5.0 | 3.0 | 2.3 |
| | | Max | -35 | +22 | -20 | -15 |
| | | Spearman | +0.912 | +0.971 | +0.982 | +0.992 |
| | FIS/04 | Average | 3.7 | 2.7 | 1.9 | 1.5 |
| | | Max | -20 | -15 | +8 | -7 |
| | | Spearman | +0.985 | +0.992 | +0.996 | +0.998 |
| Chemistry | CHIM/10 | Average | 8.5 | 4.9 | 3.8 | 2.5 |
| | | Max | -34 | +28 | -25 | -20 |
| | | Spearman | +0.921 | +0.965 | +0.979 | +0.988 |
| | CHIM/06 | Average | 4.7 | 3.1 | 2.2 | 1.4 |
| | | Max | -27 | -15 | -15 | -10 |
| | | Spearman | +0.976 | +0.989 | +0.994 | +0.997 |
| Earth Sciences | GEO/05 | Average | 17.6 | 8.6 | 4.5 | 3.5 |
| | | Max | -92 | -65 | -41 | -39 |
| | | Spearman | +0.780 | +0.922 | +0.973 | +0.983 |
| | GEO/12 | Average | 3.3 | 6.5 | 4.7 | 3.7 |
| | | Max | ±14 | -28 | -21 | ±14 |
| | | Spearman | +0.981 | +0.932 | +0.964 | +0.971 |
| Biology | BIO/03 | Average | 13.3 | 7.7 | 5.4 | 4.0 |
| | | Max | -51 | -31 | -20 | -20 |
| | | Spearman | +0.887 | +0.953 | +0.973 | +0.985 |

---

[11] The SDSs GEO/12 and MED/34 present a very low number of observations (15 and 13 respectively). Because of that, even one shift only can notably impact the average value of the whole SDS distribution.

[12] MAT/01 = Mathematical logic; MAT/07 = Mathematical physics; FIS/06 = Physics for earth and atmospheric sciences; FIS/04 = Nuclear and subnuclear Physics; CHIM/10 = Food chemistry; CHIM/06 = Organic chemistry; GEO/05 = Applied geology; GEO/12 = Oceanography and atmospheric physics; BIO/03 = Environmental and applied botanics; BIO/11 = Molecular biology; MED/34 = Physical and rehabilitation medicine; MED/03 = Medical genetics; VET/08 = Clinical veterinary medicine; AGR/05 = Forestry and silviculture; ICAR/07 = Geotechnics; ICAR/02 = Maritime hydraulic construction and hydrology; ING-IND/12 = Mechanical and thermal measuring systems; ING-INF/02 = Electromagnetic fields.

| UDAs | SDSs[12] | Index | 2004-2008 | 2005-2008 | 2006-2008 | 2007-2008 |
|---|---|---|---|---|---|---|
| | BIO/11 | Average | 4.3 | 2.8 | 2.0 | 1.5 |
| | | Max | -23 | -18 | -14 | -9 |
| | | Spearman | +0.978 | +0.988 | +0.994 | +0.997 |
| Medicine | MED/34 | Average | 22.7 | 5.1 | 5.0 | 6.2 |
| | | Max | -100 | +17 | +17 | +25 |
| | | Spearman | +0.526 | +0.966 | +0.955 | +0.943 |
| | MED/03 | Average | 3.4 | 2.3 | 1.6 | 1.1 |
| | | Max | +14 | +7 | +8 | ±5 |
| | | Spearman | +0.987 | +0.995 | +0.997 | +0.998 |
| Agricultural and veterinary sciences | VET/08 | Average | 21.8 | 10.4 | 8.1 | 5.0 |
| | | Max | -73 | -58 | -39 | -27 |
| | | Spearman | +0.754 | +0.926 | +0.935 | +0.972 |
| | AGR/05 | Average | 5.4 | 1.6 | 2.5 | 1.8 |
| | | Max | -27 | ±7 | -13 | ±7 |
| | | Spearman | +0.956 | +0.994 | +0.988 | +0.994 |
| Civil Engineering | ICAR/07 | Average | 20.0 | 7.7 | 5.6 | 3.6 |
| | | Max | -84 | -73 | -58 | -58 |
| | | Spearman | +0.729 | +0.910 | +0.955 | +0.974 |
| | ICAR/02 | Average | 8.8 | 5.3 | 3.8 | 2.3 |
| | | Max | -60 | -59 | -59 | -38 |
| | | Spearman | +0.927 | +0.960 | +0.968 | +0.991 |
| Industrial and information engineering | ING-IND/12 | Average | 24.1 | 15.0 | 9.5 | 7.1 |
| | | Max | -100 | -71 | -28 | -23 |
| | | Spearman | +0.512 | +0.787 | +0.921 | +0.954 |
| | ING-INF/02 | Average | 6.9 | 5.0 | 3.4 | 2.9 |
| | | Max | -39 | -28 | -14 | -20 |
| | | Spearman | 0.947 | 0.972 | 0.987 | 0.988 |

*Table 5: Descriptive statistics of distributions of differences of researchers' productivity rankings (percentiles) by UDA*

These results on individual researchers seem to provide only partial confirmation of a previous study (Abramo et al., 2011b), which showed that rankings lists for university productivity were relatively variable only for the disciplines of the Mathematics and engineering UDA.

In the next section we explore the question further by analyzing the extent of average shift when university rankings are given as quartile classes, rather than given as percentile listings.

### 3.3 Volatility of productivity quartile rankings of researchers

In most real-world assessment exercises performance profile of universities are expressed in quartiles, so we classified Italian researchers into four classes according to their productivity, assigning values of 4, 3, 2 and 1, corresponding to the first, second, third and fourth quartiles for the productivity distribution in the SSDs. The analysis examines the same five scenarios of different years of observation, with 2008 as benchmark. Table 6 presents the number of researchers showing one quartile variation of bibliometric productivity by UDA. Data for the 2004-2008 differences reflect various aspects of what has already emerged: Mathematics and computer sciences has the greatest percentage of researchers that make at least one shift in class (31.33%); Physics has the minimum value (15.12%). For 2007 data in this UDA, there are less than 5% (4.74%) of researchers who make the one class shift. For the same year, there are higher percentages in Agricultural and veterinary sciences (10.43%) e Earth sciences (9.35%).

| UDA | 2004-2008 | 2005-2008 | 2006-2008 | 2007-2008 |
|---|---|---|---|---|
| Mathematics and computer sciences | 537 (31.33%) | 371 (21.65%) | 247 (14.41%) | 131 (7.64%) |
| Physics | 281 (15.12%) | 186 (10.01%) | 131 (7.05%) | 88 (4.74%) |
| Chemistry | 518 (21.57%) | 304 (12.66%) | 221 (9.20%) | 163 (6.79%) |
| Earth sciences | 199 (28.19%) | 140 (19.83%) | 94 (13.31%) | 66 (9.35%) |
| Biology | 774 (22.92%) | 483 (14.30%) | 328 (9.71%) | 232 (6.87%) |
| Medicine | 1266 (21.25%) | 781 (13.11%) | 560 (9.40%) | 371 (6.23%) |
| Agricultural and veterinary sciences | 448 (30.73%) | 285 (19.55%) | 223 (15.29%) | 152 (10.43%) |
| Civil engineering | 144 (24.62%) | 109 (18.63%) | 76 (12.99%) | 43 (7.35%) |
| Industrial and information engineering | 674 (26.16%) | 476 (18.48%) | 352 (13.66%) | 190 (7.38%) |

*Table 6: Number of researchers showing quartile variations of bibliometric productivity by UDA.*

A further interesting aspect concerns the number of outliers, or researchers with shifts of two or three quartiles in productivity rank[13]. Table 7 shows that such anomalies are not excessive. Comparing rankings from citations counted in 2004 to those for benchmark 2008, only 293 researchers show two or three quartile variations. In detail, 99 of these cases fall in Industrial and information engineering; 43 are in Civil engineering, the UDA that also has the highest percentage of outliers among researchers (7.36%). In Chemistry, from 2005 onwards there are nil researchers with shifts of two or three classes; the same occurs for Physics and Earth sciences from 2006 onward.

| UDA | 2004 vs. 2008 | 2005 vs. 2008 | 2006 vs. 2008 | 2007 vs. 2008 |
|---|---|---|---|---|
| Mathematics and computer sciences | 49 (2.86%) | 3 (0.18%) | 2 (0.12%) | 1 (0.06%) |
| Physics | 11 (0.59%) | 4 (0.22%) | 0 | 0 |
| Chemistry | 6 (0.25%) | 0 | 0 | 0 |
| Earth sciences | 0 | 3 (0.42%) | 0 | 0 |
| Biology | 10 (0.30%) | 0 | 1 (0.03%) | 0 |
| Medicine | 24 (0.40%) | 3 (0.05%) | 1 (0.02%) | 0 |
| Agricultural and veterinary sciences | 51 (3.50%) | 7 (0.48%) | 1 (0.07%) | 1 (0.07%) |
| Civil engineering | 43 (7.36%) | 14 (2.40%) | 5 (0.86%) | 1 (0.17%) |
| Industrial and information engineering | 99 (3.85%) | 28 (1.09%) | 10 (0.39)% | 2 (0.08%) |

*Table 7: Number (percentage), by UDA, of researchers showing two or three quartiles variations of bibliometric productivity*

The authors also investigated the final aspect of variability in rank for researchers that place among the first or last of the standings. We define "top scientists" as those that place above the 80th national percentile for productivity in a given SDS and year. We applied a probit regression model to evaluate variation in probability for a researcher to be ranked "top" in 2008, given the rank of the same researcher in 2004. Two dummy variables were constructed: dependent variable $Y_{08}$ and independent variable $X_{04}$, which respectively take the value of 1 if the researcher is top scientist in 2008 or 2004, otherwise 0. Estimating a probit model for every UDA, we obtain the following results (Table 8). The nine models are statistically significant (p-value < 0.000) and the pseudo $R^2$ confirms a good fit. In Physics, the probability of being a top scientist in 2008 for those ranked "top" in 2004 is equal to 0.907. For this UDA the test confirms the low volatility in rankings previously seen with the analysis of quartiles. Lower probabilities are instead seen in Earth Science (0.780) and in Mathematics and Computer sciences (0.791). However, even though there is some variation among the UDAs, the probabilities seen are all quite high and consistently never lower than 0.780.

---
[13] The extreme case of a shift of three quartiles is possible if a researcher who reaches the first class in the benchmark year places in last class in a previous year, or vice versa.

| UDA | Coeff | St. Err. | $Pr(Y_{08}=1/X_{04})$ | Pseudo $R^2$ | Log likelihood |
|---|---|---|---|---|---|
| Mathematics and computer sciences | 2.510 | 0.992 | *0.791* | 0.500** | -431.76 |
| Physics | 3.304 | 0.114 | *0.907* | 0.698** | -282.03 |
| Chemistry | 2.875 | 0.889 | *0.859* | 0.591** | -493.33 |
| Earth sciences | 2.413 | 0.145 | *0.780* | 0.470** | -189.99 |
| Biology | 2.893 | 0.751 | *0.861* | 0.597** | -685.38 |
| Medicine | 3.133 | 0.060 | *0.889* | 0.658** | -1027.89 |
| Agricultural and veterinary sciences | 2.519 | 0.103 | *0.801* | 0.500** | -373.24 |
| Civil engineering | 2.660 | 0.169 | *0.828* | 0.536** | -138.43 |
| Industrial and information engineering | 2.571 | 0.078 | *0.810* | 0.513** | -638.44 |

*Table 8*: *Probit model estimates of probability of being a "top scientist" in the 2008 ranking list for researchers that are top scientists in 2004 rankings, by UDA*
*\*\* p-value < 0.000*

The results confirm that the evaluation of researchers by citation counts taken soon after the period of publication is affected by variable rates of inaccuracy across UDAs but is particularly negligible for the UDAs of Physics, Biology and Medicine. Further, the possibility for researchers to be ranked as top scientists at five years from the triennium of observation of their scientific production is strongly dependent on their rank immediately after the same triennium.

## 4. Conclusions

The advantages of bibliometric techniques for large scale evaluation could be hampered by the time necessary for citations to accumulate and represent an accurate proxy of the impact of research activity. It is understood that there is a tradeoff between timeliness of evaluation execution and accuracy of researcher performance rankings.

To provide a measure of this trade-off, we elaborated 2001-2003 productivity rankings of researchers for each SDS, for the citation windows ending in each year from 2004 to 2008. The 2008 ranking list, being the most accurate, is the benchmark for estimating inaccuracy in the rankings calculated in previous years.

The elaborations indicate that these rankings converge linearly on the benchmark, though with differences among the UDAs and among the SDSs of each single UDA: this variation is clearly as expected, given the variability in citation patterns among fields and disciplines.

In terms of correlation, the Spearman index comparing the 2008 ranking with those from previous years is never below 0.5. As soon as 2004, the correlation is actually greater than 0.9 in 117 out of the 174 total SDSs. The distribution of average values of differences in rankings also shows substantial convergence between the 2004 and 2008 rankings. In 76.4% of SDSs, the average value of shift in rank is less than or equal to 12 percentile places. The differences are particularly negligible for the discipline areas of Physics, Biology and Medicine. In general, the maximum shifts in rank are in a negative sense: positive shifts are more numerous but average less in extent. Finally, analyzing only the top scientists, we see that the possibility that scientists are at the top of their SDSs five years after the observation triennium is strongly dependent on their position in the rankings calculated immediately after the same triennium.

Thus, in line with results from previous studies, the evaluation of a researcher by means of citations received immediately after the period of publication is affected by a rate of inaccuracy that varies across disciplines but is not greatly significant, particularly in the Physics, Biology and Medicine. The accuracy of bibliometric assessment for individual scientists' productivity seems quite acceptable even immediately after a given three-year period and would clearly be even greater for observation periods longer than three years, as typically practiced in national research assessments.

A further aspect for investigation is precisely the question of the optimal length for the period of observation of research activity. In other words, once the moment for conducting an evaluation is fixed (and thus the date for observing citations), the question is how many years of observation are necessary for the rank of a researcher to stabilize? The authors intend to examine the question soon.